\documentclass[pra, onecolumn]{revtex4}
\usepackage{graphicx}

\usepackage{ulem}
\usepackage[section]{placeins}
\usepackage{amsmath}
\usepackage{amssymb}
\usepackage{dsfont}
\usepackage{natbib}

\usepackage{subfiles}
\usepackage{graphicx}
\usepackage{array}
\usepackage{calrsfs}
\usepackage{hhline}
\usepackage{relsize}

\begin{document}

\title{Intensity-corrected 4D light-in-flight imaging}

\author{Imogen Morland$^{1}$, Feng Zhu$^{1}$, Germ\'an Mora Mart\'in$^{2}$, Istvan Gyongy$^2$ and  Jonathan Leach$^{1,*}$}
\address{$^1$Institute of Photonics and Quantum Sciences, Heriot-Watt University, David Brewster Building, Edinburgh EH14 4AS, UK}
\address{$^2$Institute for Integrated Micro and Nano Systems, The University of Edinburgh, Edinburgh, EH9 3JL, UK}
\address{$^*$j.leach@hw.ac.uk}

%\date{\today}

\begin{abstract}

 Light-in-flight (LIF) imaging is the measurement and reconstruction of light’s path as it moves and interacts with objects.  It is well known that relativistic effects can result in apparent velocities that differ significantly from the speed of light.  However, less well known is that Rayleigh scattering and the effects of imaging optics can lead to observed intensities changing by several orders of magnitude along light’s path. We develop a model that enables us to correct for all of these effects, thus we can accurately invert the observed data and reconstruct the true intensity-corrected optical path of a laser pulse as it travels in air. We demonstrate the validity of our model by observing the photon arrival time and intensity distribution obtained from single-photon avalanche detector (SPAD) array data for a laser pulse propagating towards and away from the camera. We can then reconstruct the true intensity-corrected path of the light in four dimensions (three spatial dimensions and time). 
   
\end{abstract}
\maketitle

\section{\label{sec:Intro}Introduction}

\begin{figure}
\centering
\centerline{\includegraphics{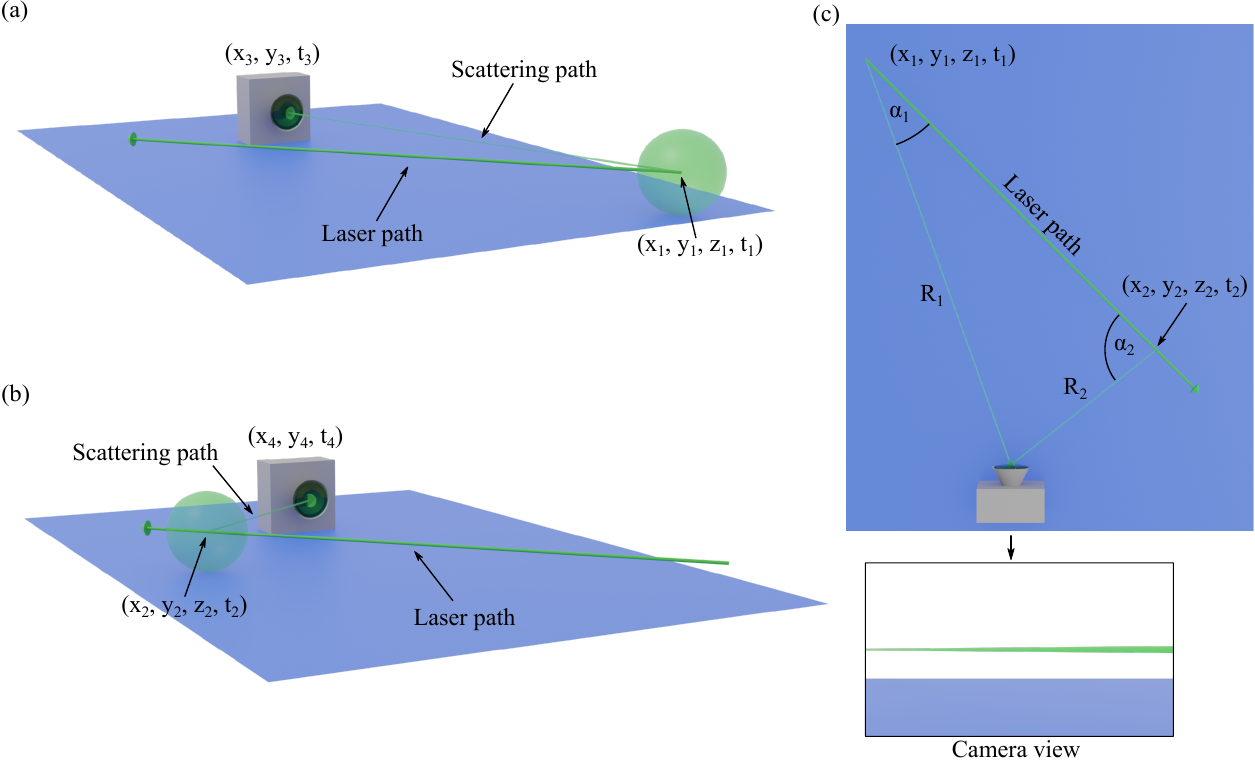}}
\caption{(a) Light in ``real space" is scattered at ($x_1, y_1, z_1, t_1$) in all directions.  A proportion of this scattered light travels to the camera, and is recorded in ``camera space" as a signal at ($x_3, y_3, t_3$), where $x_3$ and $y_3$ are pixel positions. (b) The remaining light travels across the field-of-view, and is scattered at position and time ($x_2, y_2, z_2, t_2$).  This event is recorded at ($x_4, y_4, t_4$). (c) Birds-eye-view of the two scattering events where $R_1$ and $R_2$ are the distance between the camera and the first and second scattering events respectively, and $\alpha_1$ and $\alpha_2$ are the scattering angles.  The time difference for the two events in ``camera space" is  $t_4 - t_3 = \Delta t + (R_2-R_1)/c$, whereas the time difference in ``real space" is $t_2 - t_1 = \Delta t$. Rayleigh scattering effects observed by the camera are dependent on $\alpha_1, \alpha_2, R_1$ and $R_2$, which differ significantly for the two scattering events shown. Focusing effects from the imaging optics also contribute to the intensity signal. An image rendered from the perspective of the camera shows the right side of the beam, which is closer to the camera, is larger compared to the left side of the beam.  This corresponds to a lower energy density on the right hand side, and therefore a brighter image on the left.
\label{Fig: Intro}}
\end{figure}

As light travels and interacts with objects, photons are scattered in all directions. Light-in-flight (LIF) imaging is the process of capturing scattered photons using detectors with high temporal resolution such that light's path can be reconstructed.  Three-dimensional LIF imaging was first captured using a holographic plate to record the spherical wavefronts of pulses reflected by mirrors; the technique involved no mechanical processes and achieved a temporal resolution of 800 ps \cite{abramson1978light}. This demonstrated real-time imaging of light undergoing dynamic processes, previous imaging was static and time averaged \cite{abramson1983light,abramson1989single}. Further work proposed a mechanism for correcting distortion affects when imaging light  \cite{abramson1984light}. Recent 3D LIF holography techniques use a scattering medium and achieve higher temporal resolutions \cite{hausler1996observation,kubota2007moving}.

The field of light-in-flight imaging was recently revolutionised by Velten {\it et al.} \cite{velten2013femto}, who imaged femtosecond laser pulses propagating through a scattering media using a streak camera.  This new LIF method allowed for scattering light dynamics to be observed at unprecedented temporal and spatial resolutions. However, this method requires a scanning mechanism to build a 2D image resulting in an acquisition time of one hour. Other methods for capturing 3D LIF involve transient imaging using a photonic mixer device (PMD), achieving a nanosecond temporal resolution with a one minute acquisition time \cite{heide2013low}. In addition, other 3D LIF imaging methods include time encoded amplified imaging and computer tomography, which achieve nanosecond and picosecond temporal resolutions respectively \cite{goda2009serial,li2014single}.

The type of scattering that is observed is dependent on the medium that the light propagates through. For example, light has been captured propagating through fibre optics \cite{warburton2017observation} and heated rubidium vapor \cite{wilson2017slow}. When light travels through air, Rayleigh scattering is the dominant effect, and this was captured by Gariepy {\it el al.}~who demonstrated three-dimensional LIF imaging using a single-photon avalanche detector (SPAD) array camera \cite{gariepy2015single}.  In this work, the light propagated in one plane that was perpendicular to the axis normal to the detector.  Following this, it was recognised that relativistic effects, where the apparent velocity of light would deviate from $c$, could be observed with LIF \cite{laurenzis2016relativistic, clerici2016observation} and these principles have allowed four-dimensional LIF reconstruction to be demonstrated for multiple paths of light over a narrow range of angles \cite{zheng2020computational}. This was generalised, using a megapixel camera and machine learning techniques, to capture 4D LIF imaging of multiple pulses following arbitrary paths in space \cite{Morimoto2021}.

These technologies have already been used in fluorescence lifetime imaging \cite{li2010real}, light detection and ranging (LIDAR) imaging through scattering media \cite{kocak2008focus} and imaging around corners \cite{velten2012recovering, gariepy2016detection, chan2017non}. Ultimately, the ability to accurately capture the full scattering dynamics of light could lead to new approaches when imaging deep inside the human body, see Ref.~\cite{faccio2018trillion} for an overview of light-in-flight research. Recent research tackling the problem of imaging in highly scattering media has shown computational imaging approaches can provide images in two  \cite{Lyons2019TOFtomography} and three dimensions \cite{Lindell2020tomography}.

Our work builds on the recent LIF research by developing a model to compensate for distortions in the the recorded intensity, as well as relativistic effects previously observed, and reconstruct the 4D path of laser pulses. To do this, it is necessary to understand the underlying physics of light scattering in air and the relationship to imaging optics. This is illustrated in Fig.~\ref{Fig: Intro}, where light scattered at a time $t_1$ from an object at a location ($x_1, y_1, z_1$) propagates a distance $R_1$ to a camera. The remaining light continues to propagate to position ($x_2, y_2, z_2$) where another scattering event occurs at time $t_2$, and the scattered light travels a distance $R_2$ to the camera. The total time taken for the pulse to travel between the two scattering events is $t_2 - t_1 = \Delta t$. Whereas, the two scattering events are recorded by the camera at times $t_3$ and $t_4$ respectively, and the difference in arrival time recorded by the camera is $t_4 - t_3 = \Delta t + (R_2-R_1)/c$. This means the arrival time data recorded by the camera is different to the true propagation times of light and is ultimately dependent on the propagation angle. For the case of a camera, there is a mapping of an event occurring in three spatial dimensions and time to a camera with two spatial dimensions and time. The third spatial ($z$) dimension is collapsed and contained within the temporal data of the camera.

Rayleigh scattering effects are also observed by the camera and are dependent on the scattering angles ($\alpha_1$ and $\alpha_2$) and propagation distances ($R_1$ and $R_2$). These variables vary along light's path and so the intensity contribution is dependent on the position along the path. Furthermore, focusing effects contribute to the recorded intensity profile and are dependent on the perpendicular distances between the scattering event and camera. This is shown in  the camera view image in Fig. \ref{Fig: Intro} (c) which depicts an integrated image of the laser pulse's path across the field-of-view of the camera used to render the image. The depth of field is increased such that the whole path is in focus. The pulse, travelling towards the camera from left to right, is further away from the camera on the left-hand side and is therefore focused to a smaller size than the right-hand side of the pulse. This results in the intensity of the pulse increasing as the distance between the camera and pulse increases. 

In this work, we are able to measure and subsequently correct all of the effects mentioned above.  That is to say, we can correct both the temporal distortion, arising from relativistic effects, and the intensity distortions, resulting from Rayleigh scattering and the imaging optics.
We demonstrate intensity-corrected light-in-flight imaging using a SPAD array, recording data for a laser pulse propagating at large and small angles with respect to the observation axis of the camera. The relativistic effects result in apparent speed-of-light velocities that span several orders of magnitude, and the intensity effects lead to observed intensities changing by at least a factor of two along the pulse's path.

\section{\label{sec:Theory}Theory}

\begin{figure}
\centering
\centerline{\includegraphics{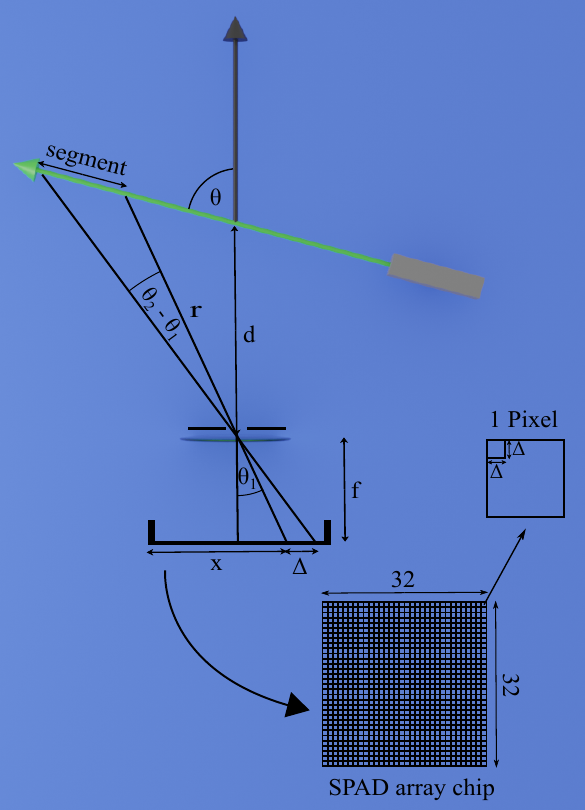}}
\caption{Theory Schematic: The intensity of a segment of the pulse is dependent on several factors: Rayleigh scattering, focusing effects and the integrated pulse length. These intensity contributions are derived using the variables shown above where $\theta$ is the propagation angle of the pulse relative to the optical axis, d is the distance between the centre of the pulse and imaging lens, f is the focal length of the imaging lens, r is the distance between the imaging lens and the nearest edge of the segment, $\theta_{1}$ is given by Equ.~(\ref{eq: theta1}), $\theta_{2}$ is given by Equ.~(\ref{eq: theta2}), x is the distance along the SPAD array to a given pixel and $\Delta$ is the active pixel width. Combining these effects, the intensity of scattered photons along the beam in camera space $(I(x; ~\theta, ~A, ~f, ~\Delta))$ is derived and given in Equ.~(\ref{eq: intensity}). The relativistic effects are explained using the same variables by Equ.~(\ref{eq: time}). 
\label{Fig: Theory time}}
\end{figure}

 Consider a pulse of light that travels in three dimensions and is imaged using a camera with high temporal resolution. To develop the theoretical framework, we introduce the concept of the ``camera space" to indicate where the data is recorded and the ``real space" to indicate the three dimensional space in which the light pulse travels. It is the goal for the work to convert the camera space data to the real space as accurately as possible. The inversion of the camera space data to the real space path enables the true intensity-corrected light path to be reconstructed.

Light-in-flight data is subject to intensity and relativistic effects observed in the camera space. The intensity of scattered photons along the beam in camera space is derived by considering the intensity contribution from one segment of the beam on one pixel. The intensity of a segment of the beam is calculated using the schematic in Fig.~\ref{Fig: Theory time} where laser pulses traveling across the field-of-view, at propagation angle $\theta$ with respect to the observation axis, are imaged using a SPAD array. A proportion of the photons within each pulse are scattered by air molecules and travel through the imaging lens aperture. Different segments of the beam are imaged by different pixels within the SPAD array and the intensity contribution from each segment is dependent on focusing effects from the imaging optics ($I_{f}$), Rayleigh scattering ($I_{r}$), and integrated path length ($I_{s}$). 

The intensity of the beam in camera space $(I(\theta_{1}, ~\theta_{2}; ~\theta))$ is given by 
\begin{equation}\label{eq: Intensity not substituted}
	I(\theta_{1}, ~\theta_{2}; ~\theta) = B  I_{f}(\theta_{1};~r,~f)I_{r}(\theta_{1};~\theta,~r)I_{s}(\theta_{1},~\theta_{2};~r),
\end{equation}
where $B$ is a normalisation constant dependent on integration time and laser power, $x$ is the distance along the SPAD array to a given pixel, $A$ is the sensor width, $f$ is the focal length of the lens, $\Delta$ is the active pixel width, $\theta$ is the propagation angle relative to the observation axis and $r$ is the distance between the imaging lens and the nearest edge of the segment, $\theta_{1}$ satisfies
\begin{equation}\label{eq: theta1}
	\theta_{1}(x;~A,~f) = \tan^{-1} \Big( \frac{2x - A}{2f} \Big),
\end{equation}
and $\theta_{2}$ satisfies
\begin{equation}\label{eq: theta2}
	\theta_{2}(x;~A,~f,~\Delta) = \tan^{-1} \Big( \frac{2x - A +2\Delta}{2f} \Big).
\end{equation}

The first contribution to the intensity of a segment of the beam is from focusing effects in the imaging optics of the system and is given by 
\begin{equation}\label{eq: focus}
	I_{f}(\theta_{1};~r,~f) = \frac{r\cos\theta_{1}}{f}.
\end{equation}
This contribution is a result of parts of the beam which are further away from the lens focusing to a smaller point on the SPAD array with higher energy density. 

The second contribution is from photons undergoing Rayleigh scattering with air molecules and is given by
\begin{equation}\label{eq: RS}
	I_{r}(\theta_{1};~\theta,~r) = \frac{I_{0} \pi^{4}(n^{2}-1)^2 d_{r}^{6}}{8 \lambda^{4} (n^{2}+2)^2 } \frac{1+\cos^2(\theta - \theta_{1})}{r^2},
\end{equation}
where $I_{0}$ is the intensity constant, n is the refractive index, $d_{r}$ is the scattering particle diameter and $\lambda$ is the wavelength of scattered light. Rayleigh scattering is dependent on the scattering angle and distance between the SPAD array and the pulse, which both change along the beam.

The final contribution to the intensity of one pixel is from the integrated path length, which is the segment length imaged by each pixel, given by
\begin{equation}\label{eq: Integrated path length}
	I_{s}(\theta_{1},~\theta_{2};~r) = \frac{r\sin(\theta_{2}-\theta_{1})}{\sin(\theta - \theta_{1})}.
\end{equation}
This results in pixels at the edge of the SPAD array seeing a larger length of pulse than pixels in the middle of the SPAD array. 

By combining these effects and substituting Equ. (\ref{eq: theta1})-(\ref{eq: Integrated path length}) into Equ. (\ref{eq: Intensity not substituted}) the intensity of scattered photons along the beam recorded by the SPAD array ($I(x; ~\theta, ~A, ~f, ~\Delta)$)is found to be
\begin{equation}\label{eq: intensity}
	I(x; ~\theta, ~A, ~f, ~\Delta) =   C\frac{1 + \cos^2 (\theta- \tan^{-1} (\frac{2x-A}{2f}))}{f \sin\theta - \frac{2x-A}{2}\cos\theta} \Big( \tan^{-1}\Big(\frac{2x-A+ 2\Delta}{2f}\Big) - \tan^{-1}\Big(\frac{2x-A}{2f}\Big) \Big),
\end{equation}
where C is a normalisation constant which includes the Rayleigh scattering constants, the integration time of the SPAD array and the optical power of the laser. Equation (\ref{eq: intensity}) assumes $\sin(\theta_{2}-\theta_{1}) \approx \theta_{2}-\theta_{1}$.

Finally, the Rayleigh effect is shown by measuring the central pixel intensity ($I_{c}(\theta;~f, ~\Delta)$) for different values of $\theta$. This intensity is independent of focusing effects as d is constant for all $\theta$ and is given by
\begin{equation}\label{eq: Intensity RS}
	I_{c}(\theta;~f, ~\Delta)  = I(x=\frac{A}{2}; ~\theta, ~A, ~f, ~\Delta) =\frac{C(1 + \cos^2 \theta)}{f\sin \theta}\tan^{-1}\left(\frac{\Delta}{f} \right) \propto  \frac{1 + \cos^2 \theta}{\sin \theta},
\end{equation}
which is derived by substituting $x = A/2$ into Equ. (\ref{eq: intensity}). This equation is a modified version of the Rayleigh scattering effect and introduces a normalisation factor that takes into account the length of pulse imaged by the central pixel.

Relativistic effects seen in the camera space result in the pulse appearing to travel at apparent velocities different to the speed of light. The arrival time in camera space is dependent on $\theta$ and d as shown in Fig. \ref{Fig: Theory time} (a). The arrival time difference between the central pixel and an arbitrary pixel ($\Delta t(x; ~\theta, ~A, ~f)$) is given by 
\begin{equation}\label{eq: time}
	\Delta t(x; ~\theta, ~A, ~f) =   \frac{d\Big( \sqrt{\big((\frac{2x-A}{2f})^{2}+1\big)}\sin\theta - \frac{2x-A}{2f} \Big)}{c(\sin\theta + \frac{2x-A}{2f}\cos\theta) }     -\frac{d}{c},
\end{equation}
where c is the speed of light in air. From the above equations, the relativistic and intensity effects observed in the camera space can be modelled and compared to experimental data.

\section{\label{sec:Set up}Experimental Set-up}

\begin{figure}
\centering
\centerline{\includegraphics{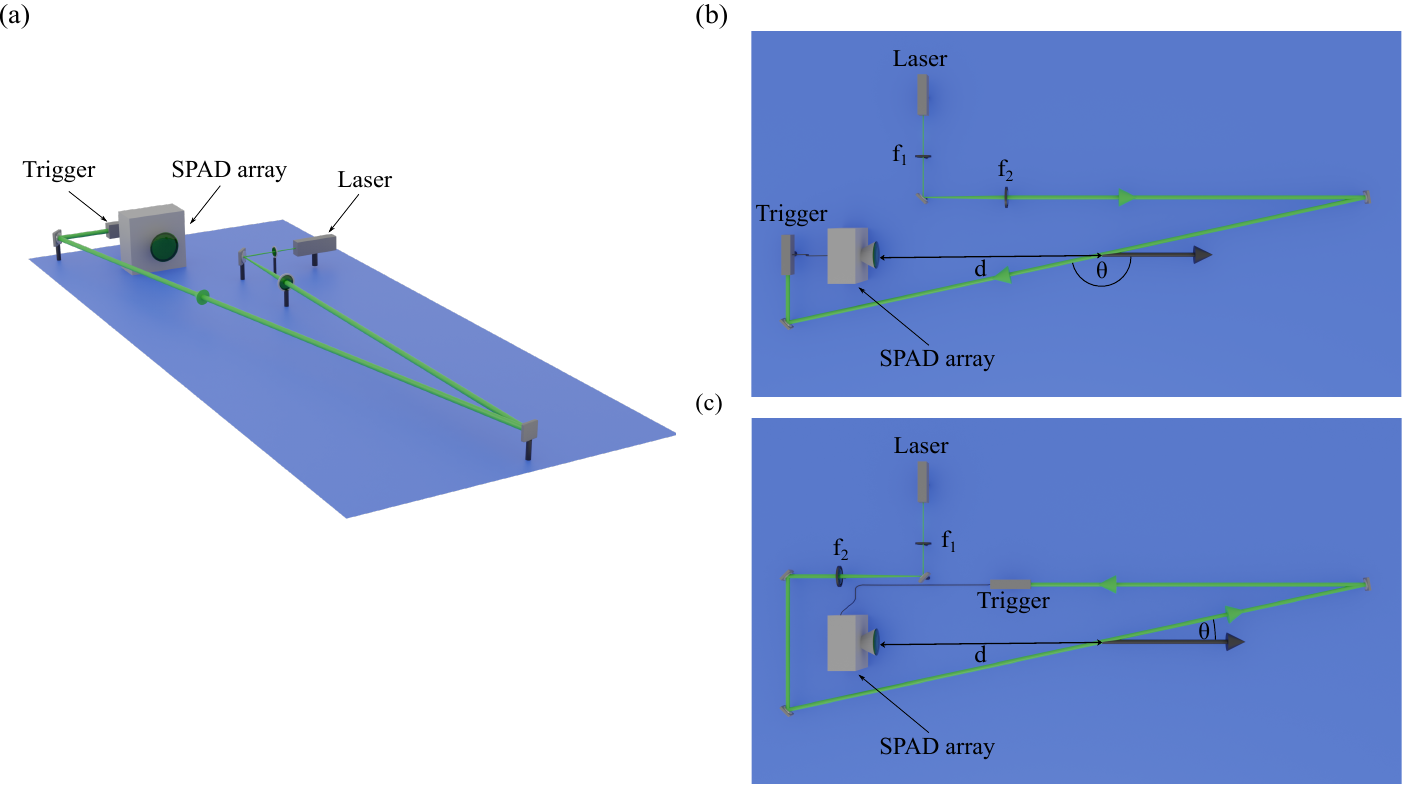}}
\caption{(a) 532 nm laser pulses are collimated by a series of lenses, increasing the beam diameter by four times to $\approx 5~$mm, and travel towards the SPAD array which records temporal and intensity data of scattered photons. From this information, $\theta$ and the intensity distribution of the beam $(I(x; ~\theta, ~A, ~f, ~\Delta))$ in camera space are calculated. (b) Birds-eye-view of pulses travelling towards the SPAD array where $d = 25.8~cm$, $f_{1}=100~ mm$ and $f_{2}=400 ~mm$. (c) Birds-eye-view of pulses travelling away from the SPAD array.
\label{Fig: set-up BASIL}}
\end{figure}

\begin{figure}
\centering
\centerline{\includegraphics{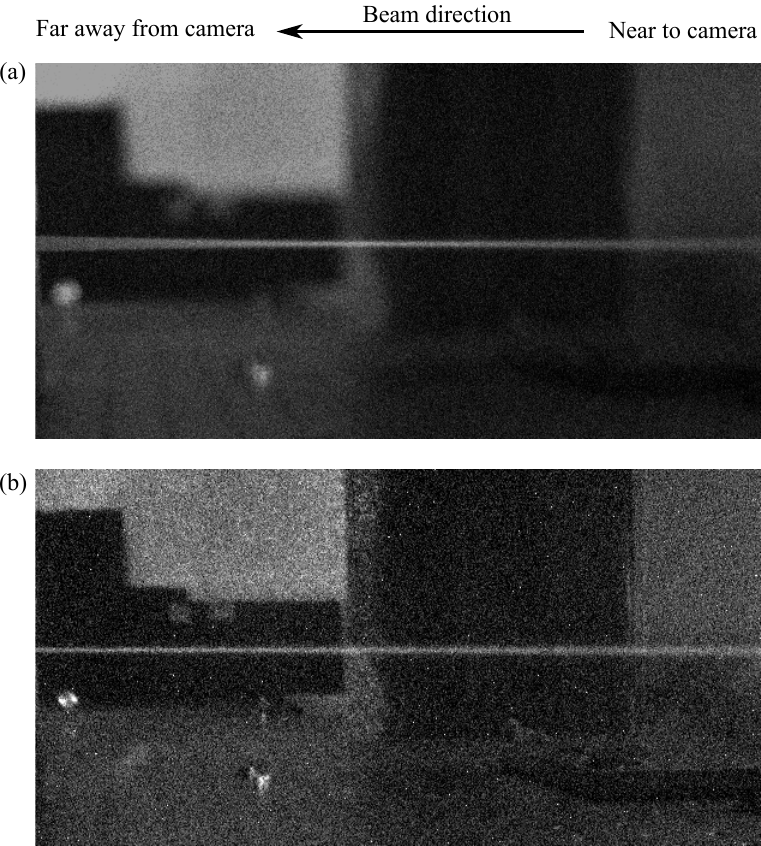}}
\caption{The effect of stopping down the aperture on the camera lens as measured with an EMCCD camera. The light is travelling away from the sensor from right to left in the images. (a) EMCCD intensity image of the beam with the aperture fully open. (b) EMCCD intensity image of the beam with the aperture closed. In (b) the entire beam is in focus and the intensity effects described by Equ.~(\ref{eq: intensity}). Our theoretical model assumes that we stop down the aperture, as seen in (b).}
\label{Fig: set-up EMCCD}
\end{figure}
The relativistic and intensity effects of LIF imaging are investigated using the experimental set-up shown in Fig.~\ref{Fig: set-up BASIL}. The system includes a SPAD array camera, a 532 nm short pulsed laser (Teem Photonics STG-03E-1x0), and an optical constant fraction discriminator used as a trigger. The impact of intensity and relativistic effects are more pronounced when the light travels at large or small angles with respect to the optical axis of the camera, which corresponds to light travelling towards and away from the camera, see Fig.~\ref{Fig: set-up BASIL} (b) and (c) respectively.

Laser pulses, which have a pulse width of $\approx$ 500 ps, are expanded to a beam waist of $\approx 5~$mm and collimated via two lenses of focal length 100 mm and 400 mm respectively, resulting in a Rayleigh range of 150 m. This ensures there are no intensity effects due to the beam diverging as it travels across the field-of-view of the sensor. The laser pulses are directed to a constant fraction discriminator acting as a trigger with 200 ps jitter, which sends 4 kHz transistor–transistor logic (TTL) pulses to the SPAD array. The TTL pulse starts the timer for each of the 32 x 32 pixels operated in time-Correlated Single Photon Counting (TCSPC) mode. Histograms of photon counts are recorded for every pixel over 1024 time bins, each with a width of 55 ps. The pixel area and active area are $50~\mu$m x $50~\mu$m and $6.95~\mu$m x $6.95~\mu$m respectively, giving a fill factor of 1.9~\%. 

An 8 mm focal length c-mount lens is used to image the beam onto the sensor. The aperture of the lens can be stopped down to extend the depth of field, and this is essential to reduce blurring and ensure the entire path of the beam is in focus on the camera. The mirrors used to direct the beam towards the SPAD array are placed outside the field-of-view so only photons scattered by air molecules are collected by the imaging lens, thus avoiding saturation effects and allowing Rayleigh scattering to be observed. Finally, to observe the Rayleigh scattering effects, the laser and trigger were placed on a rotation stage system which allows $\theta$ to be easily varied.

When measuring $I(x; ~\theta, ~A, ~f, ~\Delta)$, it is important for the whole of the beam to be in focus. This is demonstrated in Fig.~\ref{Fig: set-up EMCCD} (a) and (b) which shows electron multiplying charge-coupled device (EMCCD) intensity images of the beam travelling from right to left away from the SPAD array with an open and closed aperture respectively. When the lens aperture is open, out of focus light contributes to the intensity image resulting in part of the beam being out of focus and less intense than predicted by Equ.~(\ref{eq: intensity}). When the lens aperture is closed, only in focus light is incident on the detector and the intensity effects predicted are observed. This condition requires longer acquisition times to collect sufficient photon counts to build an intensity image. 

\section{\label{sec:Result}Results}

%%%BASIL
%Towards
%%Camera space
\begin{figure}
\centering
\centerline{\includegraphics{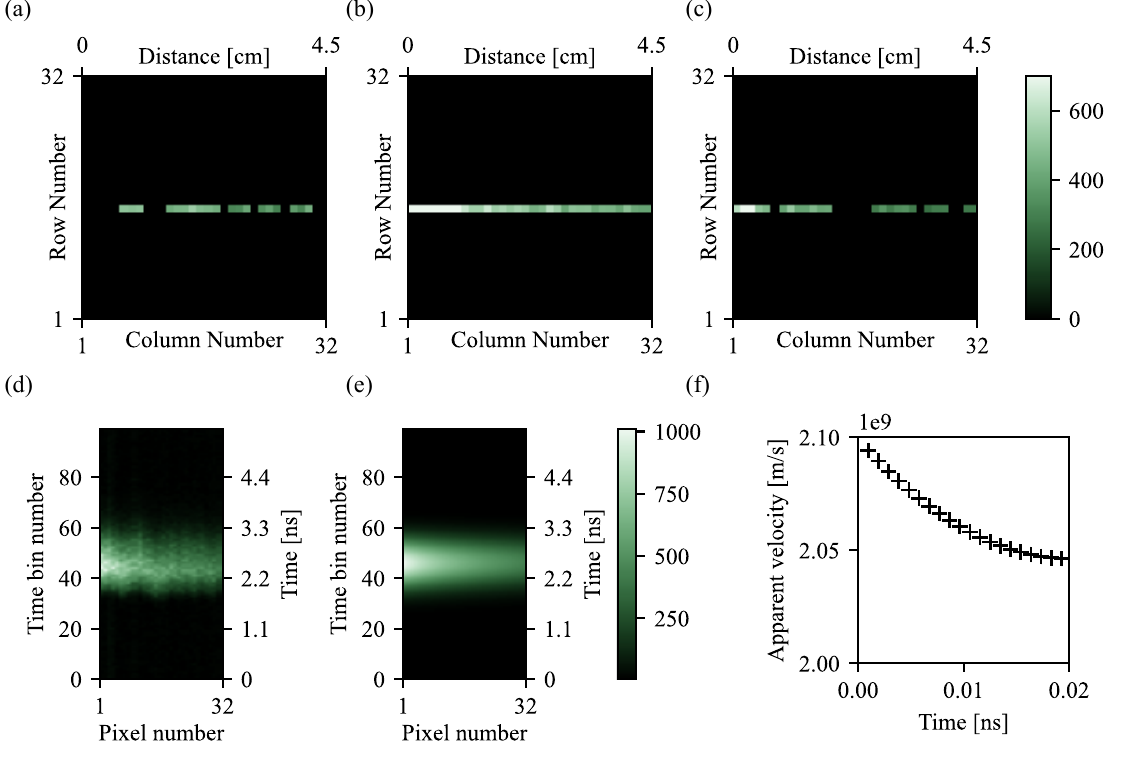}}
\caption{Results Camera Space: $\theta=167.0^{\circ}\pm0.5^{\circ}$: laser pulses travelling towards the SPAD array using the set-up show in Fig. \ref{Fig: set-up BASIL} (a). (a)-(c) Three frames of laser pulses travelling from left to right across row 17 are shown at 2.1 ns, 2.5 ns and 2.8 ns, where the colour bar represents the number of photon counts and the distance axis is the horizontal field-of-view in real space. (d)-(e) Data and fitted model for row 17 of the SPAD camera. This shows photon counts as a function of position and time. The theoretical fit to the data calculates $\theta$ as $165.5^{\circ}\pm0.1^{\circ}$. (f) The apparent velocity of the pulse as a function of propagation time varies from 7.0 c to 6.8 c.  Note that the timescale is the camera space time.  This is significantly shorter than the real space time, leading to the apparent superluminal velocities. 
\label{Fig: Result BASIL towards}}
\end{figure}

In order to achieve intensity-corrected 4D LIF imaging, it is important to remove the noise present in the SPAD array data. This has been achieved by fitting Gaussian functions to each pixel and setting the pixel intensity to zero if the standard deviation of the Gaussian is outside an acceptable range. Furthermore, for noisy pixels within the beam path, interpolation is performed. 

Camera space results for laser pulses travelling from left to right towards the SPAD array at $\theta=167.0^{\circ}\pm0.5^{\circ}$ are shown in Fig. \ref{Fig: Result BASIL towards}. Three frames, each with time duration 55 ps, of laser pulses travelling across row 17 of the SPAD array are shown at 2.1 ns, 2.5 ns and 2.8 ns in Fig. \ref{Fig: Result BASIL towards} (a) to (c). The time taken for the pulse to travel across the SPAD array is less than the time bin width, resulting in the pulse being present for all pixels of row 17 in a single frame.

The data and fitted model for row 17 as a function of position and time is given in Fig. \ref{Fig: Result BASIL towards} (d) and (e).   The photon intensity decreases as pixel number increases in both the data and the fitted model.   This is because the left hand side of the beam is further away from the SPAD array and so focuses to a smaller area on the detector with a higher energy density.

The total time taken for the pulse to travel across the SPAD array was measured to be 18 ps, and using equations (\ref{eq: time}) and (\ref{eq: intensity}) to fit a 2D Gaussian function to the data, $\theta$ was calculated to be $165.5^{\circ}\pm0.1^{\circ}$. The error in $\theta$ was calculated by numerically creating 10 statistically identical data sets and fitting to these.  These additional data sets were created by sampling from a Poisson distribution with a mean and variance determined by the initial experimental data.  Finally, the apparent velocity of the pulse as a function of propagation time is shown in Fig. \ref{Fig: Result BASIL towards} (f) and varies from 6.8 c to 7.0 c.  This superluminal apparent velocity is entirely due to the pulse travelling toward the camera.

\begin{figure}
\centering
\centerline{\includegraphics{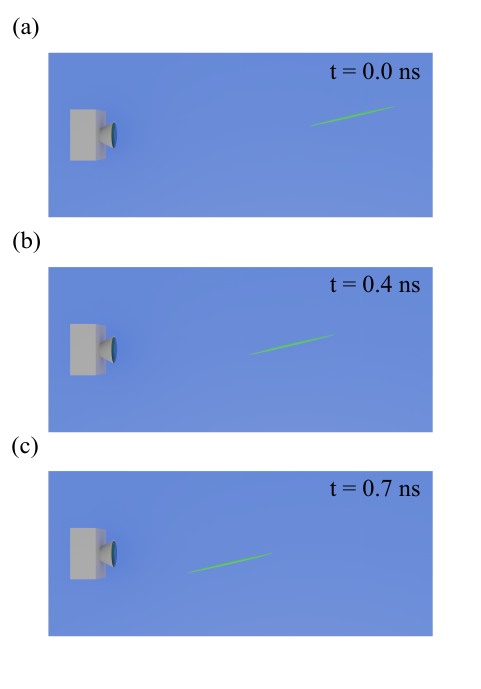}}
\caption{Results Real Space: $\theta=167.0^{\circ}\pm0.5^{\circ}$: Three frames of the pulse traveling towards the SPAD array at $165.5^{\circ}\pm0.1^{\circ}$. The pulse position is calculated using $\theta$ and the fitted intensity data is used to normalise the pulse intensity.  Note that the timescale is now real space time as the pulse of light travels at c. %The beam diameter was measured as $\approx 5~$mm and the pulse width is known to be 15 cm.
\label{Fig: Result Real BASIL towards}}
\end{figure}

%%Real space
Using the data obtained by the SPAD array, the camera space data is converted to real space data using $\theta$ and normalising the data by the fitted intensity values at the appropriate times. Three frames from the real space movie of the pulse traveling towards the SPAD array are shown in Fig. \ref{Fig: Result Real BASIL towards} (a) to (c) at 0.0 ns, 0.4 ns and 0.7 ns. The beam diameter used for the real space reconstruction was $5~$mm and the pulse width was 15 cm.  These values were taken from measurements and known values of the pulse.
 
%%AWAY
%%Camera space

\begin{figure}
\centering
\centerline{\includegraphics{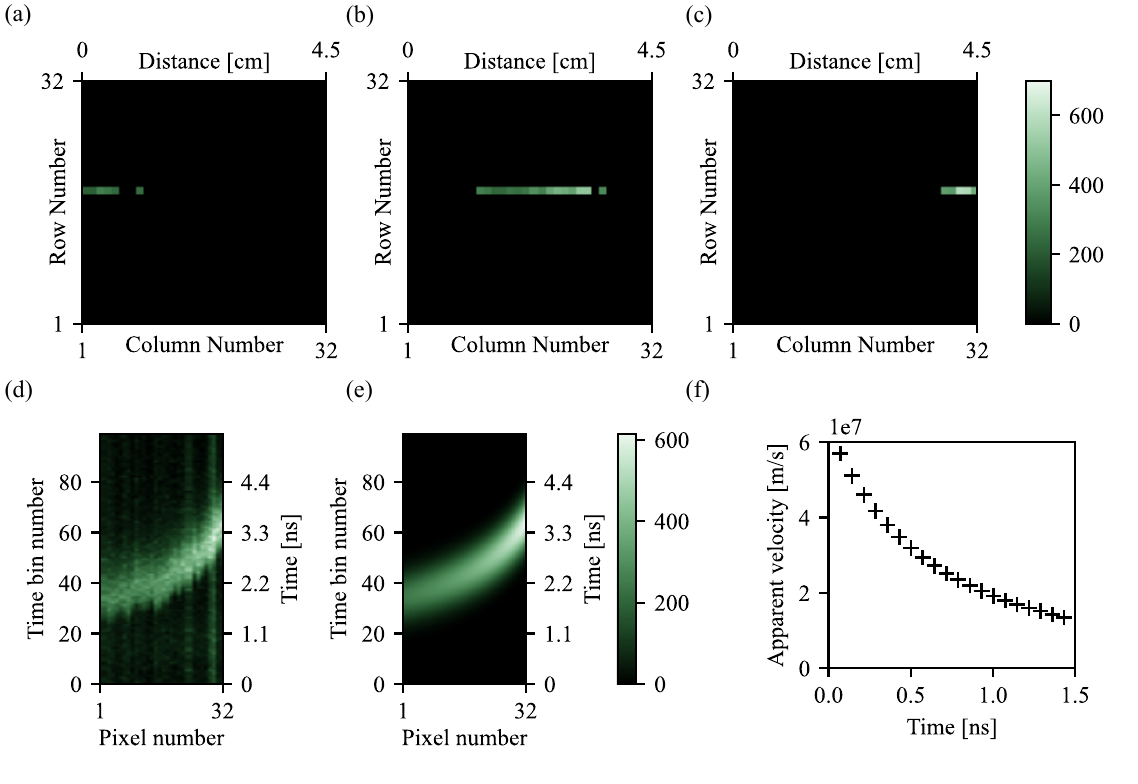}}
\caption{Results Camera Space: $\theta=13.0^{\circ}\pm0.5^{\circ}$: laser pulses travelling away from the SPAD array using the set-up show in Fig. \ref{Fig: set-up BASIL} (b). The pulse direction has been reversed for clarity. (a)-(c) Three frames of laser pulses travelling across row 17 of the SPAD array are shown at 1.7 ns, 2.5 ns and 3.3 ns. (d)-(e) Data and fitted model for row 17 of the SPAD camera. This shows photon counts as a function of position and time.  The theoretical fit to the data calculates $\theta$, as $13.9^{\circ}\pm0.1^{\circ}$. (f) The apparent velocity of the pulse varies from 0.19 c to 0.05 c, indicating a decelerating pulse travelling away from the SPAD array.  Note that the timescale is the camera space time.  This is longer than the real space time, leading to the apparent subluminal velocities.  
\label{Fig: Result BASIL away}}
\end{figure}

Camera space data was also recorded for laser pulses travelling from left to right away from the SPAD array at $\theta=13.0^{\circ}\pm0.5^{\circ}$ using the set-up shown in Fig. \ref{Fig: set-up BASIL} (b). Three frames of laser pulses travelling across row 17 are shown at 1.7 ns, 2.5 ns and 3.3 ns in Fig. \ref{Fig: Result BASIL away} (a)-(c). The pulse length present in each frame is shorter for light travelling away from the SPAD array, indicating lower apparent velocities. 

The data and model used to calculate $\theta$ are shown in Fig. \ref{Fig: Result BASIL away} (d) and (e) respectively. The total time taken for light to travel in camera space is 1.4 ns, and the curvature of the fitted function indicates the pulses appear to decelerate as they travel away from the SPAD array. Using Equ. (\ref{eq: time}) and (\ref{eq: intensity}) to fit to the data, $\theta$ was estimated to be $13.9^{\circ}\pm0.1^{\circ}$.

Finally, the apparent velocity of the pulse as a function of propagation time is shown in Fig. \ref{Fig: Result BASIL away} (f) and varies from 0.19 c to 0.05 c. This results in a ratio of the fastest to slowest apparent velocities equal to 156.  This is the largest ratio of super to subluminal apparent velocities in 4D LIF imaging; the previous highest ratio was 17, reported in Ref.~\cite{Morimoto2021}. 

\begin{figure}
\centering
\centerline{\includegraphics{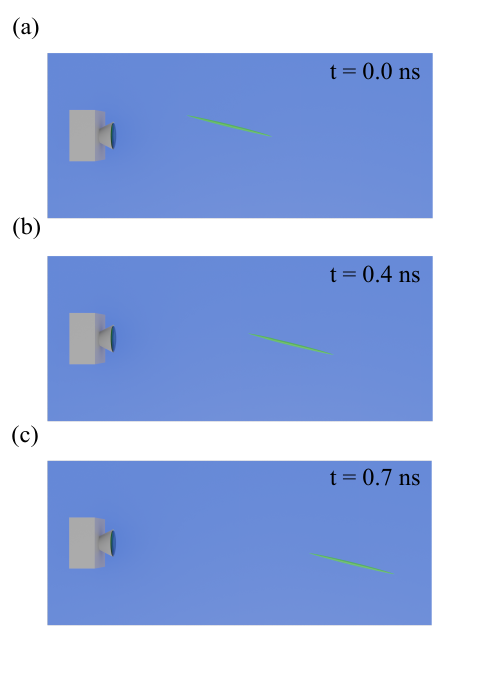}}
\caption{Results Real Space $\theta=13.0^{\circ}\pm0.5^{\circ}$: Three frames from the real space movie of the pulse traveling at $13.9^{\circ}\pm0.1^{\circ}$.  Note that the timescale is now real space time as the pulse of light travels at c.
\label{Fig: Result Real BASIL away}}
\end{figure}

Using the camera space data in Fig. \ref{Fig: Result BASIL away},  we can then recreate the real space data of the pulse travelling away from the SPAD array. Three frames from a movie showing the real space are shown in Fig. \ref{Fig: Result Real BASIL away} at times of 0.0 ns, 0.4 ns and 0.7 ns.

%RS

\begin{figure}
\centering
\centerline{\includegraphics{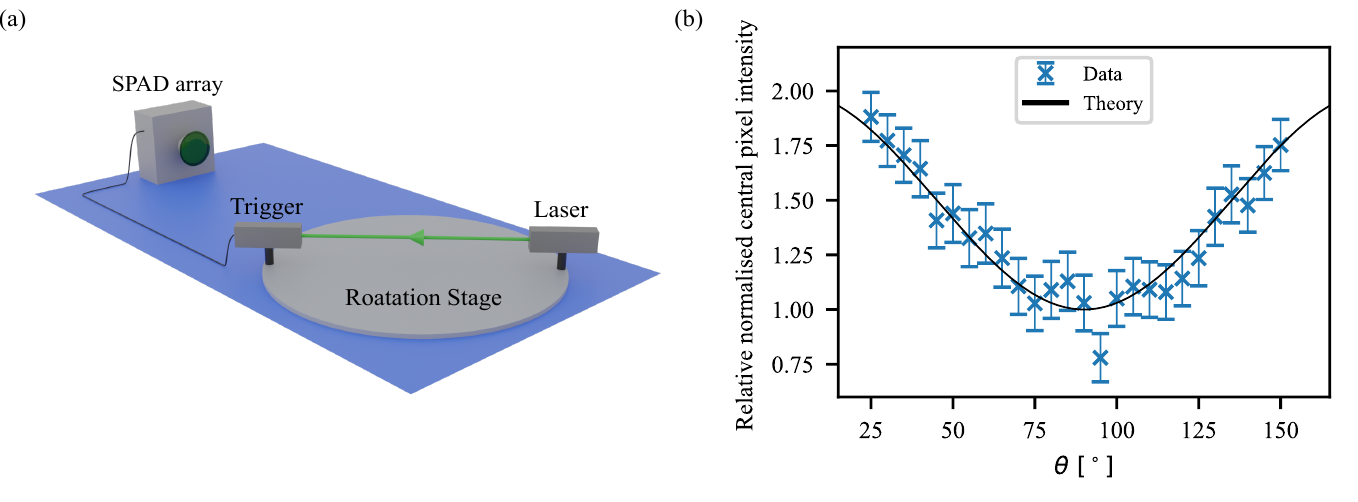}}
\caption{Experiment setup and results for Rayleigh scattering.  (a) Laser pulses travel across the SPAD array field-of-view at an angle $\theta$ set by the rotation stage. Temporal and intensity data is recorded in $5^{\circ}$ intervals between  $25^{\circ}$ and $150^{\circ}$. (b) The normalised central pixel intensity ($I_{c}(\theta;~f, ~\Delta)$) versus $\theta$. The errors are given by the square root of the total number of photons recorded $\sqrt n$. \label{Fig: Result RS}}
\end{figure}

Our final experiment demonstrates the angle dependence of scattering in air for light-in-flight imaging, i.e.~Rayleigh scattering.  This is achieved by placing the pulsed laser on a rotation stage, see Fig.~\ref{Fig: Result RS}, allowing $\theta$ to be easily altered, and recording the intensity of the central pixel  of the camera. The central pixel intensity is only dependent on $\theta$ as the distance between the centre of the rotation stage and SPAD array is constant for all values of $\theta$. This removes the effects of focusing and the inverse square dependence, which are both present in the first experiment. Fig.~\ref{Fig: Result RS} (b) shows the observed experiment data in good agreement with the predictions of Rayleigh scattering, see Equ.~(\ref{eq: Intensity RS}).    It should be noted that the effects of Rayleigh scattering were present in the previous experimental results but were harder to isolate.  

\section{\label{sec:Conclusion}Conclusion}

Relativistic effects, focusing, and Rayleigh scattering all play a significant role in the observed signal for light-in-flight imaging. By modelling these effects we have been able to invert SPAD array data and reconstruct the true 4D path of laser pulses, showing a strong agreement between experiment and theory.   We demonstrate the validity of our model by fitting to data obtained for light travelling towards and away from a SPAD array and comparing the temporal and intensity distributions to the model. The ratio of the apparent velocity of the pulses travelling towards and away is over two orders of magnitude and is the highest ratio observed for light-in-flight imaging.

\section*{Funding}
Engineering and Physical Science Research Council (EP/T00097X/1 and EP/S001638/1).

\section*{Disclosures}

The authors declare no conflicts of interest.


\begin{thebibliography}{10}
\newcommand{\enquote}[1]{``#1''}
\providecommand\JournalTitle[1]{#1}



  \bibitem{abramson1978light}
N.~Abramson, \enquote{Light-in-flight recording by holography,}
  {\protect\JournalTitle{Optics letters}} \textbf{3}, 121--123 (1978).

\bibitem{abramson1983light}
N.~Abramson, \enquote{Light-in-flight recording: high-speed holographic motion
  pictures of ultrafast phenomena,} {\protect\JournalTitle{Applied optics}}
  \textbf{22}, 215--232 (1983).

\bibitem{abramson1989single}
N.~H. Abramson and K.~G. Spears, \enquote{Single pulse light-in-flight
  recording by holography,} {\protect\JournalTitle{Applied optics}}
  \textbf{28}, 1834--1841 (1989).

\bibitem{abramson1984light}
N.~Abramson, \enquote{Light-in-flight recording. 3: Compensation for optical
  relativistic effects,} {\protect\JournalTitle{Applied optics}} \textbf{23},
  4007--4014 (1984).

\bibitem{hausler1996observation}
G.~H{\"a}usler, J.~Herrmann, R.~Kummer, and M.~Lindner, \enquote{Observation of
  light propagation in volume scatterers with 10 11-fold slow motion,}
  {\protect\JournalTitle{Optics Letters}} \textbf{21}, 1087--1089 (1996).

\bibitem{kubota2007moving}
T.~Kubota, K.~Komai, M.~Yamagiwa, and Y.~Awatsuji, \enquote{Moving picture
  recording and observation of three-dimensional image of femtosecond light
  pulse propagation,} {\protect\JournalTitle{Optics express}} \textbf{15},
  14348--14354 (2007).
  

\bibitem{velten2013femto}
A.~Velten, D.~Wu, A.~Jarabo, B.~Masia, C.~Barsi, C.~Joshi, E.~Lawson,
  M.~Bawendi, D.~Gutierrez, and R.~Raskar, \enquote{Femto-photography:
  capturing and visualizing the propagation of light,}
  {\protect\JournalTitle{ACM Transactions on Graphics (ToG)}} \textbf{32}, 1--8
  (2013).

\bibitem{heide2013low}
F.~Heide, M.~B. Hullin, J.~Gregson, and W.~Heidrich, \enquote{Low-budget
  transient imaging using photonic mixer devices,} {\protect\JournalTitle{ACM
  Transactions on Graphics (ToG)}} \textbf{32}, 1--10 (2013).

\bibitem{goda2009serial}
K.~Goda, K.~Tsia, and B.~Jalali, \enquote{Serial time-encoded amplified imaging
  for real-time observation of fast dynamic phenomena,}
  {\protect\JournalTitle{Nature}} \textbf{458}, 1145--1149 (2009).

\bibitem{li2014single}
Z.~Li, R.~Zgadzaj, X.~Wang, Y.-Y. Chang, and M.~C. Downer, \enquote{Single-shot
  tomographic movies of evolving light-velocity objects,}
  {\protect\JournalTitle{Nature communications}} \textbf{5}, 1--12 (2014).

\bibitem{warburton2017observation}
R.~Warburton, C.~Aniculaesei, M.~Clerici, Y.~Altmann, G.~Gariepy, R.~McCracken,
  D.~Reid, S.~McLaughlin, M.~Petrovich, J.~Hayes \emph{et~al.},
  \enquote{Observation of laser pulse propagation in optical fibers with a spad
  camera,} {\protect\JournalTitle{Scientific reports}} \textbf{7}, 43302
  (2017).

\bibitem{wilson2017slow}
K.~Wilson, B.~Little, G.~Gariepy, R.~Henderson, J.~Howell, and D.~Faccio,
  \enquote{Slow light in flight imaging,} {\protect\JournalTitle{Physical
  Review A}} \textbf{95}, 023830 (2017).

\bibitem{gariepy2015single}
G.~Gariepy, N.~Krstaji{\'c}, R.~Henderson, C.~Li, R.~R. Thomson, G.~S. Buller,
  B.~Heshmat, R.~Raskar, J.~Leach, and D.~Faccio, \enquote{Single-photon
  sensitive light-in-fight imaging,} {\protect\JournalTitle{Nature
  communications}} \textbf{6}, 1--7 (2015).

\bibitem{laurenzis2016relativistic}
M.~Laurenzis, J.~Klein, and E.~Bacher, \enquote{Relativistic effects in imaging
  of light in flight with arbitrary paths,} {\protect\JournalTitle{Optics
  letters}} \textbf{41}, 2001--2004 (2016).

\bibitem{clerici2016observation}
M.~Clerici, G.~C. Spalding, R.~Warburton, A.~Lyons, C.~Aniculaesei, J.~M.
  Richards, J.~Leach, R.~Henderson, and D.~Faccio, \enquote{Observation of
  image pair creation and annihilation from superluminal scattering sources,}
  {\protect\JournalTitle{Science advances}} \textbf{2}, e1501691 (2016).

\bibitem{zheng2020computational}
Y.~Zheng, M.-J. Sun, Z.-G. Wang, and D.~Faccio, \enquote{Computational 4d
  imaging of light-in-flight with relativistic effects,}
  {\protect\JournalTitle{Photonics Research}} \textbf{8}, 1072--1078 (2020).

\bibitem{Morimoto2021}
K.~Morimoto, M.-L. Wu, A.~Ardelean, and E.~Charbon, \enquote{Superluminal
  motion-assisted four-dimensional light-in-flight imaging,}
  {\protect\JournalTitle{Phys. Rev. X}} \textbf{11}, 011005 (2021).

\bibitem{li2010real}
D.-U. Li, J.~Arlt, J.~Richardson, R.~Walker, A.~Buts, D.~Stoppa, E.~Charbon,
  and R.~Henderson, \enquote{Real-time fluorescence lifetime imaging system
  with a 32$\times$ 32 0.13 $\mu$m cmos low dark-count single-photon avalanche
  diode array,} {\protect\JournalTitle{Optics express}} \textbf{18},
  10257--10269 (2010).

\bibitem{kocak2008focus}
D.~M. Kocak, F.~R. Dalgleish, F.~M. Caimi, and Y.~Y. Schechner, \enquote{A
  focus on recent developments and trends in underwater imaging,}
  {\protect\JournalTitle{Marine Technology Society Journal}} \textbf{42},
  52--67 (2008).

\bibitem{velten2012recovering}
A.~Velten, T.~Willwacher, O.~Gupta, A.~Veeraraghavan, M.~G. Bawendi, and
  R.~Raskar, \enquote{Recovering three-dimensional shape around a corner using
  ultrafast time-of-flight imaging,} {\protect\JournalTitle{Nature
  communications}} \textbf{3}, 1--8 (2012).

\bibitem{gariepy2016detection}
G.~Gariepy, F.~Tonolini, R.~Henderson, J.~Leach, and D.~Faccio,
  \enquote{Detection and tracking of moving objects hidden from view,}
  {\protect\JournalTitle{Nature Photonics}} \textbf{10}, 23 (2016).

\bibitem{chan2017non}
S.~Chan, R.~E. Warburton, G.~Gariepy, J.~Leach, and D.~Faccio,
  \enquote{Non-line-of-sight tracking of people at long range,}
  {\protect\JournalTitle{Optics express}} \textbf{25}, 10109--10117 (2017).

\bibitem{faccio2018trillion}
D.~Faccio and A.~Velten, \enquote{A trillion frames per second: the techniques
  and applications of light-in-flight photography,}
  {\protect\JournalTitle{Reports on Progress in Physics}} \textbf{81}, 105901
  (2018).

\bibitem{Lyons2019TOFtomography}
A.~Lyons, F.~Tonolini, A.~Boccolini, A.~Repetti, R.~Henderson, Y.~Wiaux, and
  D.~Faccio, \enquote{Computational time-of-flight diffuse optical tomography,}
  {\protect\JournalTitle{Nature Photonics}} \textbf{13}, 575--579 (2019).

\bibitem{Lindell2020tomography}
D.~B. Lindell and G.~Wetzstein, \enquote{Three-dimensional imaging through
  scattering media based on confocal diffuse tomography,}
  {\protect\JournalTitle{Nature Communications}} \textbf{11}, 4517 (2020).

\end{thebibliography}
\end{document}